# LORAX: Loss-Aware Approximations for Energy-Efficient Silicon Photonic Networks-on-Chip

Febin Sunny[1], Asif Mirza[1], Ishan Thakkar[2], Sudeep Pasricha[1] and Nikdast Mahdi[1]
[1]Colorado State University, Fort Collins, CO, USA, [2]University of Kentucky, Lexington, KY, USA
{febin.sunny, mirza.baig, sudeep, Mahdi.nikdast}@colostate.edu, igthakkar@uky.edu

## ABSTRACT
The approximate computing paradigm advocates for relaxing accuracy goals in applications to improve energy-efficiency and performance. Recently, this paradigm has been explored to improve the energy efficiency of silicon photonic networks-on-chip (PNoCs). In this paper, we propose a novel framework (*LORAX*) to enable more aggressive approximation during communication over silicon photonic links in PNoCs. Given that silicon photonic interconnects have significant power dissipation due to the laser sources that generate the wavelengths for photonic communication, our framework attempts to reduce laser power overheads while intelligently approximating communication such that application output quality is not distorted beyond an acceptable limit. To the best of our knowledge, this is the first work that considers loss-aware laser power management and multilevel signaling to enable effective data approximation and energy-efficiency in PNoCs. Simulation results show that our framework can achieve up to 31.4% lower laser power consumption and up to 12.2% better energy efficiency than the best known prior work on approximate communication with silicon photonic interconnects, for the same application output quality.

## KEYWORDS
Energy efficiency; PNoCs; multi-level signalling; Floating point approximation;

## 1. INTRODUCTION

The overall energy consumption in computing systems is increasing rapidly because of the continuous growth in data volumes consumed in emerging applications. Ensuring fault-free computing for such large quantities of data is becoming difficult due to various reasons. One is the fact that the increasing resource demands for big data processing limit the resources available for traditional redundancy-based fault tolerance; another more fundamental problem is the ongoing scaling of semiconductor devices, which makes them increasingly sensitive to variations, *e.g.*, due to imperfect fabrication processes. Approximate computing, which trades-off "acceptable errors" during execution to reduce energy and runtime, is a promising solution to both these challenges [1]. With diminishing performance-per-watt gains from Dennard scaling, leveraging such aggressive techniques to achieve energy-efficiency is becoming increasingly important.

To cope with the data processing needs of emerging applications, the core counts in manycore processors have also been rising. Such increase in the core counts in response to increasing processing load creates greater core-to-core and core-to-memory communication. Consequently, the traffic in the on-chip communication architecture fabric has been increasing to the point where today it costs more energy to retrieve and move data than to process it. Conventional electrical interconnects and electrical networks-on-chip (ENoCs) today dissipate very high power to support the high bandwidths and low latency requirements of data-driven parallel applications [2]. Fortunately, chip-scale silicon photonics has emerged in recent years as a very promising development to enhance NoCs with light speed photonic links that can overcome the bottlenecks of slow and noise-prone conventional electrical links. Silicon photonics can enable photonic networks-on-chip (PNoCs) that can sustain much higher bandwidths and lower latencies than ENoCs [3].

Typical PNoC architectures employ several photonic devices such as photonic waveguides, couplers, splitters, and multi-wavelength laser sources, along with microring resonators (MRs) as modulators, detectors, and switches. A laser source (either off-chip or on-chip) generates light with one or more wavelengths, which is coupled by an optical coupler to an on-chip photonic waveguide. This waveguide guides the input optical power of potentially multiple wavelengths (often referred to as wavelength-division-multiplexed (WDM) transmission), via a series of optical power splitters, to the individual nodes (*e.g.*, processing cores) on the chip. Each wavelength serves as a carrier for a data signal. Typically, multiple data signals are generated at a source node in the electrical domain as sequences of logical 1 and 0 voltage levels. These input electrical data signals are modulated onto the wavelengths using a bank of modulator MRs (*e.g.*, 32-bit data modulated on 32 wavelengths), using on-off keying (OOK) modulation. Once the data has been modulated on the wavelengths at the source node, it is routed over the PNoC till it reaches its destination node, where the wavelengths are coupled out of the waveguide by a bank of detector MRs, which drop the wavelengths of light onto photodetectors to recover the data in the electrical domain. Each node in the PNoC can communicate to multiple other nodes through such WDM-enabled photonic waveguides in the PNoC.

Unfortunately, light signals suffer losses as they propagate through waveguides, requiring high laser power to compensate for such losses, so that the signal can be received at the destination node with sufficient power to enable error-free recovery of the transmitted data. Power is also dissipated due to MR tuning at the source and destination MR banks, to ensure appropriate modulation and coupling of signals. Typically, however, the laser power dominates overall power in PNoCs. Novel solutions are therefore urgently needed to reduce this laser power footprint, so that PNoCs can serve as a viable high-bandwidth and low-latency network in emerging and future manycore architectures.

In this paper, we explore the use of data approximation to reduce the overall power and energy footprint of the laser power source in PNoCs. The main contributions of this paper can be summarized as follows:

- We develop an approach that relies on approximating a subset of data transfers for applications, to reduce energy consumption in PNoCs while still maintaining acceptable output quality;
- We explore the sensitivity of application output to varying data transfer approximation degrees and laser power levels;
- We propose an aggressive approximate strategy that adaptively switches between two modes of approximate data transmission, based on the photonic signal loss profile along the traversed path;
- We further evaluate the impact of utilizing multilevel signaling (pulse-amplitude modulation) instead of on-off keying signaling during approximate transfers for even greater energy efficiency;
- We evaluate our proposed framework (called *LORAX*) on multiple applications and contrast it with the best known prior work on approximating data transfers over PNoC architectures.

## 2. RELATED WORK

By carefully relaxing the requirement for computational correctness, it has been shown that many applications can execute with a much lower energy consumption, without significantly impacting application output quality. As an example, it is possible to approximate the weights (*e.g.*, from 32-bit floating point to 8-bit fixed point) in deep neural networks, with negligible changes in the output classification accuracy [5]. Beyond machine learning models, many other approximation tolerant applications exist, e.g., in the domains of video, image, and audio processing and big data analysis [6]. For such applications, approximation is an effective technique to improve energy efficiency.

Approximate computing solutions proposed to date can be broadly categorized into four types based on their scope [7]: hardware, storage, software, and systems. The approximation of hardware components allows a reduction in their complexity and thus a reduction in area and energy consumption [8] (*e.g.*, using an approximate full adder that inexactly computes the least significant bits, compared to a conventional full adder). Storage approximation utilizes techniques, such as reduced refresh rates in DRAM [9] which results in a deterioration of stored data, but at the advantage of increased energy efficiency in memory units. Software approximation includes algorithmic approximation, which may leverage domain specific knowledge [10]-[12] or simplify the implementation [13]. It may also refer to approximating annotated data, variables, and high-level programming constructs (*e.g.*, loop iterations), as specified by the software designer via annotations in the software program [4]. At the system level, approximation involves modification of architectures to support approximate operations. In general, attempts to create approximate NoC architectures to reduce the energy cost for communication at the system level (between processing cores and memories) would fall under this category.

Several efforts have attempted to approximate data transfers over electrical NoC architectures, by using strategies that reduce the number of bits or packets being transmitted, to reduce NoC utilization and thus reduce communication energy. An approximate NoC for GPUs was discussed in [13], where the authors proposed an approach for data approximation at the memory controller by coalescing packets with similar (but not necessarily the same) data, to reduce the packets that traverse over the reply network plane. A hardware data approximation framework with an online data error control mechanism for high performance NoCs was presented in [14]. The architecture facilitates approximate matching of data patterns, within a controllable value range, to compress them and thereby reducing the volume of data movement across the chip. A dual voltage NoC is proposed in [15], where the lower priority bits in a packet are transferred at a lower voltage level, which may cause them to incur bit flips. The higher priority bits of the packet, including headers, are transmitted with higher voltage, ensuring a lower bit error rate (BER) for them. This approach allows a trade-off between errors introduced due to the low transmission voltage and the subsequent increase in the BER, with low power consumption during transfers.

As for photonic NoCs, a recent paper [16] explored the use of approximate data communication on PNoCs for the first time. The authors explored different levels of laser power for transmission of bits across a single-writer-multiple-reader (SWMR) photonic waveguide, with a lower level of laser power used for bits which could be approximated, causing them to suffer higher BER. The work focused specifically on approximation of floating point data, which are known to be resilient to approximation compared to integer data. The least significant bits (LSBs) of the floating point data were subjected to lower laser power for transmission. However, the specific number of these bits to be transmitted as well as the laser power levels were decided in an application-independent manner, which ignores application-specific sensitivity to approximation. Moreover, the laser power is set statically, without considerations of varying loss that photonic signals encounter as they traverse through photonic waveguides.

The framework discussed in this paper (called *LORAX*) overcomes the limitations of [16] by utilizing a novel loss-aware approach that adapts laser power at runtime to enable efficient approximate communication in PNoCs. We perform comprehensive analysis of the impact of adaptive approximation and laser power levels on application output quality, to enable an approach that can be tuned in an application-specific manner. We also additionally explore the impact of discarding the conventional on-off keying photonic signaling approach in favor of a pulse amplitude modulation photonic signaling approach, on the energy savings achievable in PNoCs. To the best of our knowledge, this is the first work that considers loss-aware laser power management and multilevel signaling for approximation and energy-efficiency in PNoCs.

## 3. BACKGROUND: FLOATING POINT DATA FORMAT

In many applications, floating point data is resilient to at least some level of approximation. The Least Significant Bits (LSBs) are considered for approximation in [16], as well as in this work, as opposed to the Most Significant Bits (MSBs) due to the unique data representation for floating point data as per the IEEE-754 standard.

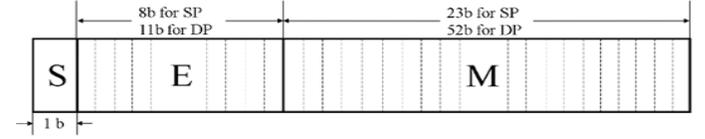

**Fig. 1: IEEE 754 floating point representation**

The IEEE-754 standard defines a standardized floating point data representation which consists of three parts: sign (S), exponent (E), and mantissa (M), as shown in Fig. 1. The true value of the data stored is:

$$X = (-1)^S \times 2^{E-bias} \times (1 + M), \quad (1)$$

where X is the floating point value. The *bias* values are 127 and 1203 for single and double precision representation respectively, and are used to ensure that the exponent is always positive, thereby eliminating the need to store the exponent sign bit. The single precision (SP) and double precision (DP) representations vary in the number of bits allotted to the exponent and mantissa (Fig. 1). E is 8 bits for SP and 11 bits for DP; while M is 23 bits for SP and 52 bits for DP. Also, S is 1 bit for both cases. From (1) we can observe how significant the S and E values are as they notably affect the value of X, but M is typically less sensitive to alterations in many cases, and it also takes up a significant portion of the floating point data representation. We consider S and E as MSBs that should not be altered, whereas M makes up the LSBs that are more suitable for approximation to save energy during photonic transmission.

We evaluate the breakdown of integer and floating point data usage across multiple applications, to establish how effective an approach that focuses on approximating floating point LSB data can be. We selected the ACCEPT benchmark suite [12], which consists of several applications that have been shown to have a relatively strong potential for approximations. We used the gem5 [22] system-level simulator and performed a benchmark characterization for this suite. We used the simulator to count the total number of integer and floating point packets in transit during the simulations. Fig. 2 shows the breakdown of the float and integer packets across the applications for large input workloads. The large input workloads were generated for applications such as *sobel* and *jpeg*, while for application from the PARSEC [23] benchmark suite, the large input workloads were selected from that suite.

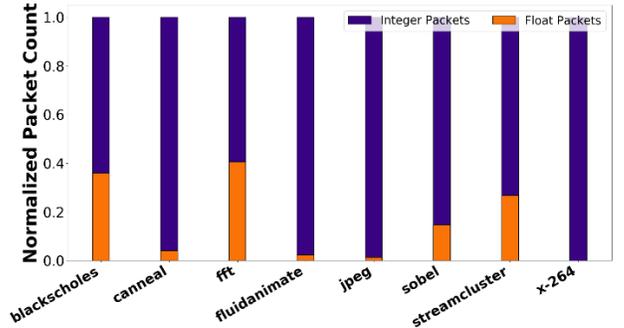

**Fig. 2: ACCEPT benchmark application characterization**

From Fig. 2 it is apparent that applications utilize varying number of floating point and integer data. To evaluate our proposed framework, we focus on five of these applications with notable and diverse floating point communication, while excluding *fluidanimate* and *x264*, owing to their negligible floating point traffic. We also selected *jpeg* as a case study into the effects of approximation on low floating point traffic data.

## 4. *LORAX* FRAMEWORK: OVERVIEW

This section discusses the components of our *LORAX* (LOss-awaRe ApproXimation) framework. Section 4.1 provides an overview of our loss-aware laser power management strategy. Section 4.2 discusses our integration of multilevel signaling to further enhance this approach.

### 4.1 Loss-aware laser power management for approximation

The laser power required at a source node to transfer data on a WDM photonic waveguide (link) to a destination node can be expressed as:

$$P_{laser} - S_{detector} \geq P_{phot\_loss} + 10 \times \log_{10} N_\lambda \quad (2)$$

where $P_{laser}$ is the laser power in dBm, $S_{detector}$ is the MR detector sensitivity (*e.g.*, -20 dBm [27]), and $N_\lambda$ is the number of wavelength channels in the link. Also, $P_{phot\_loss}$ is the photonic loss incurred by the signal in its transmission, which includes propagation and bending losses in the waveguide, through losses in MR modulators and detectors, modulating losses in modulator MRs, and detection loss in detector MRs. $P_{laser}$ thus depends on the link bandwidth in terms of $N_\lambda$, and the total loss $P_{phot\_loss}$ encountered by the photonic signals traversing the waveguide. The $P_{phot\_loss}$ encountered along the waveguide reduces the optical signal power, and the signal can only be accurately recovered at the destination node if the received signal power is higher than $S_{detector}$. Ensuring this requires a high enough $P_{laser}$ to compensate for the losses.

To approximate data transmission for floating point data transfers, [16] used lower $P_{laser}$ for transmitting LSBs (while keeping $P_{laser}$ untouched for MSBs). However, if the destination node is relatively farther along a waveguide from a source node, the signals would encounter high losses and the signal intensity at the detector MRs would be less than $S_{detector}$, which would result in detecting logic '0' for all the LSB signals at the destination node. In the scenario where the destination is closer to the source, it may be possible to detect the LSB signals accurately, as long as the losses encountered are low enough that the signal power at the detector MRs would be higher than $S_{detector}$, even with the reduced $P_{laser}$ for the LSBs.

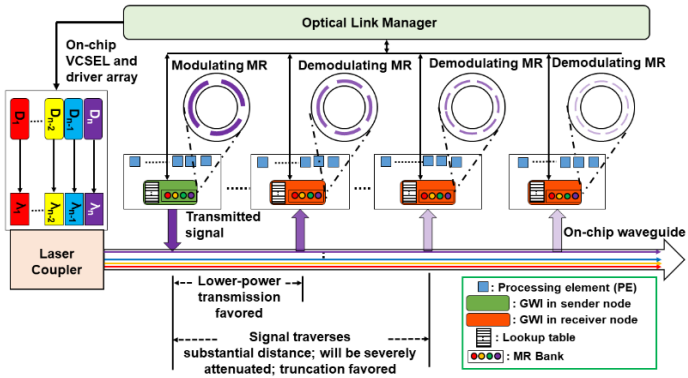

**Fig. 3: Overview of our proposed *LORAX* framework**

We make the following observation about the approach in [16]: for each communication on a waveguide, if we are aware of the distance of the destination from the source, it is possible to calculate the losses encountered for the signals, which can allow us to determine whether the signals can be recovered accurately, or if they will be detected as all '0's. In such a scenario, it is more energy-efficient to simply truncate all the LSBs (*i.e.*, reduce $P_{laser}$ to 0 for LSB signals) when the destination is farther along the waveguide and there is no likelihood of the signal being recovered accurately ([16] still advocates for sending the LSB signals at reduced $P_{laser}$ even if the signals cannot be recovered at the destination). In the cases where the destination is closer to the source, we can transmit the LSB signals with a lower $P_{laser}$, allowing some of the data be detected accurately at the destination, while approximating other data depending on its content and distance to the destination. Unlike [16] which reduces $P_{laser}$ to a fixed value for a fixed subset of the LSB signals, irrespective of the application, we conjecture that it is important to tune the appropriate number of LSB signals and $P_{laser}$ level in an application-specific manner. This is because the outputs for each application are sensitive to the LSB values in different ways, so a one size fits all approach, as proposed in [16], may not make sense.

Our proposed *LORAX* framework is motivated by the shortcomings in [16] and the observations discussed above. Fig. 3 shows the operational details of our framework on a single writer multiple reader (SWMR) waveguide that is part of a PNoC architecture. Note that while we illustrate our framework with an SWMR waveguide, our framework is also applicable (with minimal changes) to multiple writer multiple reader (MWMR) and multiple writer single reader (MWSR) waveguides that are also used in many PNoCs. In the SWMR waveguide as shown in Fig. 3, only one sender node is active per data transmission phase and one out of multiple (three in the figure) receiver nodes is the destination for the transmission. In a pre-transmission phase (called the receiver selection phase) the sender will notify the receivers about the destination for the upcoming data transmission, and only the destination node will activate its MR banks, whereas the other nodes will power down their MR banks to save power in the transmission phase. As shown in Fig. 3, if the destination node is close to the sender node (*e.g.*, the leftmost out of the three potential destination nodes), we can transmit the LSB signals with a lower $P_{laser}$ as shown in Fig. 4(b). Otherwise if the destination node is farther away from the sender node (*e.g.*, the second out of the three potential destination nodes shown in Fig. 3), we determine that it would not be possible to detect the LSB signals at that destination due to the greater losses the signals will encounter. Therefore, we dynamically turn off $P_{laser}$, essentially truncating the LSB bits, as shown in Fig. 4(a).

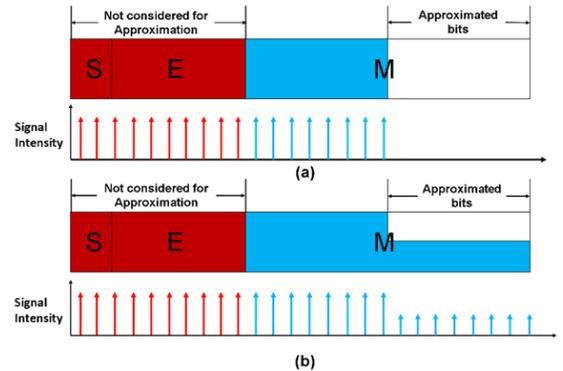

**Fig 4: LSB signal transmission (a) truncation (b) lower laser power**

To implement this framework, we require a laser control mechanism that can dynamically control the laser power being injected into the on-chip waveguides. For this, we utilize an on-chip laser array with vertical-cavity surface-emitting lasers (VCSELs) [17], which can be directly controlled using on-chip laser drivers. With the laser drivers, we can control the power fed into each individual VCSEL, thus controlling the intensity of the laser output for a particular wavelength corresponding to that VCSEL. The Gateway Interface (GWI) that connects the electrical layer of the chip to the PNoC (Fig. 3), communicates the desired $P_{laser}$ intensity level (including 0 for truncation) to the drivers, via an optical link manager, similar in structure to the one proposed in [18].

Our approach also requires each source node to know when to switch between truncation and a lower $P_{laser}$ level, and also whether the packet contains approximable data or not. Identification of candidate packets to be approximated is done at the processing element level, via source code annotations [4], to generate a flag for data (*e.g.,* floating point) that is approximable. This flag is inserted in the packet header. The GWI can then read the flag to determine if the packet is to be approximated. Then we must determine whether the approximation is to be done via reduced power transmission or truncation. This requires a lookup table at each GWI (Fig. 3), with the IDs of all the destination GWIs to which truncated transmission should be preferred. The table consists of loss values to each destination from the source. The values can be easily calculated offline and used to populate the table, as the location of destination nodes as well as the cumulative loss to their GWI from the source does not change at runtime. We discuss the overheads of the tables in Section 5.1. An application-specific $P_{laser}$ for the LSB signals, discussed further in Section 5.2, can be used to determine if the signals can be detected at the given destination GWI, by consulting the loss value to that destination from the table, and then a decision can be made to either truncate or transmit the LSB bits. Once the decision to truncate or transmit at a lower laser power is made, the required intensity levels for the wavelengths are communicated to the VCSEL drivers via the optical link manager.

### 4.2 Integrating multilevel signaling for approximation

The discussion in the previous section assumes the use of conventional on-off keying (OOK) signal modulation, where each photonic signal can have one of two power levels: high or on (when transmitting '1') and low or off (when transmitting '0'). In contrast, multilevel signaling is a signal modulation approach where more than two power levels of voltage are utilized to transmit multiple bits of data simultaneously in each photonic signal. The obvious perk with such multilevel signaling is the increased bandwidth it provides. Leveraging this technique in the photonic domain has, however, traditionally been a cumbersome process with high overheads, *e.g.*, when using the signal superposition techniques from [20]. But with advances such as the introduction of Optical Digital to Analog Converter (ODAC) circuits [21] that are much more compact and faster than Mach-Zehnder Interferometers (MZIs) used in techniques involving superimposition [20], multilevel signaling has been shown to be more energy-efficient than OOK [19], making it a promising candidate for more aggressive energy savings in photonic links.

Four-level pulse amplitude modulation (PAM4) is a multilevel signal modulation scheme where two extra levels of voltage (or photonic signal intensity) are added in between the 0 and 1 levels of OOK. This allows PAM4 to transmit 2 bits per modulation as opposed to 1 bit per modulation in OOK. This in turn increases the bandwidth when compared to OOK. While PAM4 promises better energy efficiency than OOK, it is prone to higher BER due to having multiple levels of the signal close to each other in the spectrum. Thus we cannot reduce the laser power level of the LSB bits to the level used in OOK, as it would significantly reduce the liklihood of accurate data recovery even when destination nodes are relatively close to the source. We therefore keep the reduced laser power level for PAM4 to 1.5× that of OOK. This may seem like a backward step in conserving energy, but the reduced operational cost per modulation and the reduced wavelength count for achieving the same bandwidth as OOK, may reduce the overall laser power. The experimental results in the next section quantify the impact and trade-off of using PAM4 signaling with our framework.

## 5. EXPERIMENTS
### 5.1 Experimental setup

To evaluate our framework, we consider the Clos PNoC architecture [24], with a baseline OOK signaling. The Clos PNoC (Fig. 5) has an 8-ary 3 stage topology for a 64-core system with 8 clusters and 8 cores per cluster. Inter-cluster communication utilizes the photonic waveguides in the PNoC. Each cluster has two concentrators and a group of 4 cores connected to a concentrator, where the concentrators communicate with each other via an electrical router. The PNoC architecture was modeled and simulated using a SystemC based cycle-accurate simulator. The gem5 simulator was used for full system simulation, to generate traces for the entire application that were replayed on the PNoC simulator to determine energy savings in the PNoC. Then, details of the approximate data communication (*i.e.*, whether a packet was truncated or transmitted at lower power) were used to modify data in a subsequent gem5 simulation, to estimate the impact of the approximation on output quality for the application being considered.

**Table 1: 64-core architecture configuration**

| Simulated component | Specification |
|---|---|
| No. of cores, processor type | 64, x86 |
| DRAM | 8GB, DDR3 |
| Memory controllers | 8 |
| L1 I/D cache, line size | 128KB each, direct mapped, 64B |
| L2 cache, line size, coherence | 2MB, 2-way set associative, 64B, MESI |

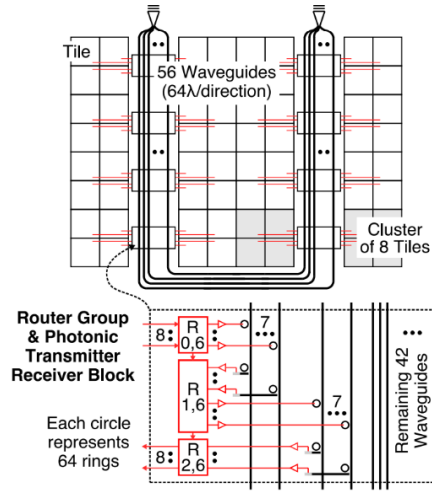

**Fig. 5: 8-ary 3 stage Clos architecture with 64 cores [24]**

**Table 2: Loss and power values considered for photonic devices**

| Parameters considered | Parameter values |
|---|---|
| Detector sensitivity | -23.4 dBm [30] |
| MR Through loss | 0.02 dB [28] |
| MR Drop Loss | 0.7 dB [32] |
| Waveguide propagation loss | 0.25 dB/cm [33] |
| Waveguide bend loss | 0.01 dB/90º [31] |
| Thermo-optic tuning | 240 µW/nm [29] |

Table 1 shows the gem5 architectural parameters considered for the platform used in our experiments. As discussed earlier, six applications from the ACCEPT benchmark were used in our evaluations. The performance was evaluated at the 22nm CMOS node for a 400$mm^2$ chip, with cores and routers operating at 5GHz clock frequency. DSENT [25] was used to calculate the energy consumption by routers and the GWI at each node. CACTI [26] was used to evaluate the power and area for the lookup tables in the GWIs. These values were found to be: 0.105 $mm^2$ of area consumption for all tables, with a total power overhead of 0.06 $mW$. A single cycle latency overhead was considered for accessing the 64-entry table at 22nm. We considered $N_\lambda = 64$ for OOK, which would enable 64 bit transmission across the waveguide per cycle. For PAM4, we only need to consider $N_\lambda = 32$ to achieve the same bandwidth as with OOK transmission. Table 2 shows the energy values for losses and power dissipation in different photonic devices. These values are used to calculate laser power from (2) and total power after considering tuning and lookup table overheads. We additionally consider a PAM-4 induced signaling loss of 5.8dB in $P_{phot\_loss}$ for laser power calculations for PAM-4. To compensate for the increased sensitivity of PAM4 to bit errors, we also consider laser power levels that are 1.5× than those used for OOK signaling. Lastly, we calculated the output error incurred by the application due to an approximation approach as:

$$Percentage\ (Output)\ Error = \frac{|approximate\ value - exact\ value|}{exact\ value} \times 100 \quad (3)$$

For our analysis, we assume an error threshold of 10% output error, *i.e.*, we want to ensure that none of the approximation strategies degrade output quality by more than 10%.

### 5.2 Application-specific approximation sensitivity analysis

Our first set of experiment involves analyzing the sensitivity of an application to varying degrees of approximation of their floating point data. We were interested in studying the impact on output error of approximating a varying number of LSBs. Additionally, we were also interested to study the impact on output error of varying levels of lowered laser power for the LSBs. Fig. 6 shows the results of our comprehensive study for the six benchmarks we considered (see Fig. 2). Each of the six surface plots presents insights into the behavior of the individual applications. The z-axis shows the percentage error (PE) in application output, as a function of the reduction in $P_{laser}$ level for the photonic signals that carry the LSB bits (x-axis; varying from 0% to 100%, where 100% refers to truncation), and the number of LSBs that were considered for approximation (y-axis; with the number of bits ranging from 4 to 32).

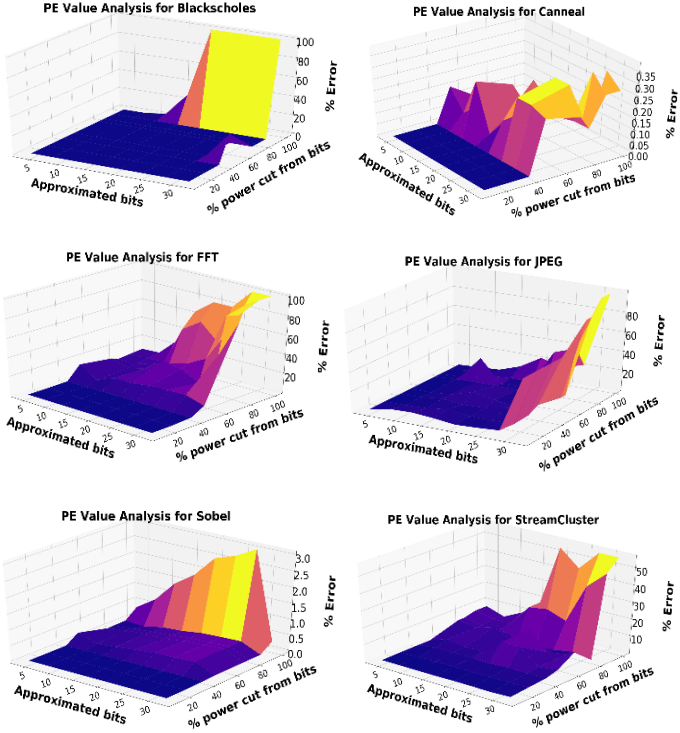

**Fig. 6:** Percentage error (PE) in application output as a function of the number of approximated LSB signals and reduction in laser power for the LSB signals, for the blackscholes, canneal, fft, jpeg, sobel, and streamcluster benchmarks with large input workloads

From the analysis it is clear that not all applications can tolerate the same level of approximation. From the PE values, we can observe that *FFT* with a large volume of floating point data traffic (see Fig. 2) reaches the error threshold of 10% rather quickly as the number of approximated bits increase and laser power levels reduce, whereas *Canneal* with a lower floating point traffic volume observed seems to have very low PE values across the various experiments (note how the z-axis only goes up to 0.35% error). The edge detection algorithm *Sobel* performs well in approximated conditions similar to *Canneal*, possibly owing to the lowered data accuracy requirements to construct the output (edges detected in an input image). *Streamcluster* involves an approximation strategy for data streams, and is also observed to be quite resilient to greater levels of approximation. *Blackscholes,* which performs market options calculations, is particularly sensitive to the approximated number of bits and the laser power levels. *JPEG* performs image compression and the output image quality is also more sensitive to approximation.

Table 3 summarizes the best combination of approximable bits (that are part of LSBs) and the laser power transmission levels for these bits, for each application, while ensuring that the application output error does not exceed 10% for our proposed framework (*LORAX*; rightmost two columns). In the next subsection, we compare *LORAX* with the framework from [16] and an approach involving truncation. Table 3 also shows the number of bits that can be truncated, selected to meet the <10% PE constraint. For the approach in [16] we perform approximation on 16 LSBs transmitted at 20% laser power (advocated as an optimal choice in that work) which also satisfies the <10% PE constraint.

**Table 3: Number of LSBs for approximation and laser transmission power level for LSB signals across benchmarks**

| Application Name | Truncation Truncated Bits | [16] | LORAX Approximated Bits | % Power reduction |
|---|---|---|---|---|
| **Blackscholes** | 12 | 16, with 20% power reduction | 32 | 90 |
| **Canneal** | 32 | | 32 | 100 |
| **FFT** | 8 | | 32 | 50 |
| **JPEG** | 20 | | 24 | 80 |
| **Sobel** | 32 | | 32 | 100 |
| **Streamcluster** | 12 | | 28 | 80 |

In Fig. 7 we use *JPEG* as an example to illustrate the effectiveness of the parameters we have chosen for it (similar analyses done for other applications is omitted for brevity). Fig. 7(a) shows the original output from the application without any approximation. Fig. 7(b) is the output when 24 LSBs of the floating point data are transmitted at 20% laser power (*i.e.,* 80% power reduction) in *LORAX*, which results in <10% PE in the output, and a relatively good output image quality. Fig. 7(c) and 7(d) show the impact of much more aggressive approximations at 28 LSBs and 32 LSBs, respectively, transmitted at 20% laser power in both cases, which shows easily observable undesirable artefacts in the output image. This serves as an example for why application specific analysis is necessary while considering approximation strategies.

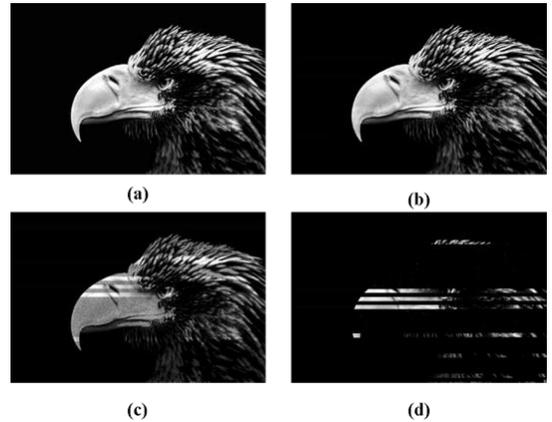

**Fig. 7:** Effects of approximation parameters on JPEG output (a) original image output from JPEG; (b) 24 LSBs approximated and 20% laser power (as in Table 3); (c) 28 LSBs approximated and 20% laser power; (d) 32 bits approximated and 20% laser power.

### 5.3 Comparative results for laser power and EPB

The analysis from the previous subsection is used to determine the application-specific laser power intensity control in our framework. We compare the laser power and energy per bit (EPB) results for two variants of our framework: with OOK (*LORAX-OOK*) and with PAM4 (*LORAX-PAM4*). We compare our two framework variants with the framework from [16] and a truncation strategy that statically truncates a fixed number of bits, with the approximated LSBs and laser power levels for our *LORAX* frameworks chosen as discussed in the previous subsection, such that output error does not exceed 10%.

Fig. 8 shows the EPB and laser power comparison results for the various frameworks on the Clos PNoC architecture. Fig. 8(a) shows that using *LORAX-OOK* results in lower EPB than [16] and the truncation approach. The truncation approach sometimes performs better than [16], as it avoids wasteful transmission at lower laser power when it is unlikely that the destination can recover the transmitted data due to high losses. But the lower number of truncated bits compared to approximated bits in [16] results in lower EPB for [16] in other cases. The *LORAX-OOK* framework improves upon both [16] and truncation, by adaptively switching between truncation and an application-specific laser power intensity level for LSBs. The *LORAX-PAM4* variant of our framework achieves the largest reduction in EPB, even though it uses higher power levels for the approximated bits. The use of fewer wavelengths in PAM4 allows for more energy savings, despite greater losses and the use of more laser power per wavelength than *LORAX-OOK*.

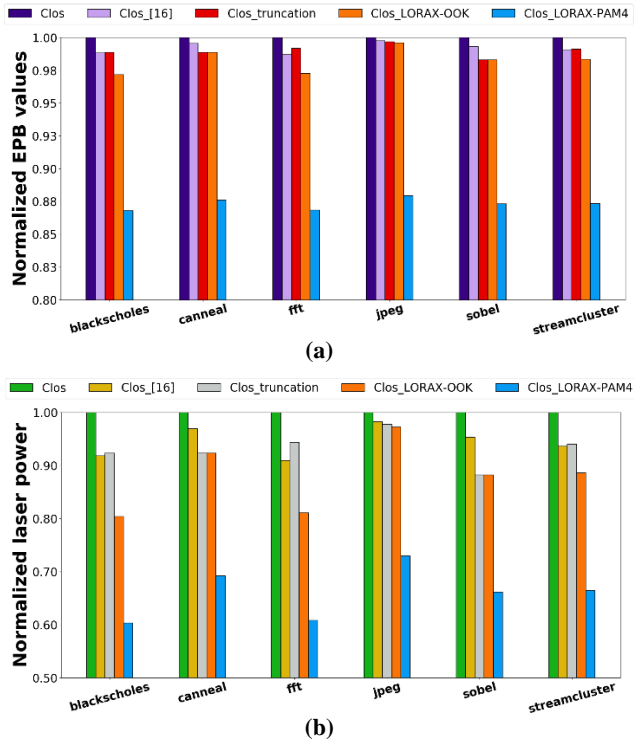

**Fig. 8: (a) Energy-per-bit (EPB) comparison across frameworks, (b) laser power comparison across frameworks**

On average, *LORAX-PAM4* shows 13.01%, 12.16%, and 12.2% lower EPB compared to the baseline Clos, [16], and truncation approaches respectively. *LORAX-OOK* exhibits 2.5%, 1.9%, and 1% lower EPB on average compared to the same approaches. In the best case scenarios for the *Blackscholes* and *FFT* applications, *LORAX-PAM4* has 13.7% and 13.5% lower EPB than the Clos baseline; and 12% and 12.2% lower EPB than [16], while against truncation it shows 12.45% and 12.4% lower EPB for these two applications.

Fig. 8(b) specifically shows the laser power reduction. On average, *LORAX-PAM4* uses 34.17%, 30.1%, and 27.2% lower laser power compared to the baseline Clos, [16], and truncation approaches respectively, while *LORAX-OOK* exhibits 12.2%, 8.1%, and 7.8% lower average laser power consumption on average. For the best case *Blackscholes* and *FFT* applications laser power for *LORAX-PAM4 is* 39.7% and 39.2% lower than the Clos baseline and 30.8% and 31.4% lower than [16], while against truncation it is 32% and 33.6% lower. These results highlight the promise of our proposed *LORAX* framework, to trade-off output correctness with energy-efficiency and laser power savings in PNoC architectures executing selected applications.

## 6. CONCLUSIONS

In this paper, we proposed a new framework called *LORAX* for loss-aware approximation of floating point data communicated over PNoC architectures in emerging manycore platforms. We also investigated how multilevel signaling can assist with the proposed approximation framework. Our results indicate that utilizing multilevel signaling as part of our framework can reduce laser power consumption by up to 39.7% over a baseline PNoC architecture. Our framework also shows up to 31.4% lower laser power and up to 12.2% better energy efficiency compared to the best known prior work on approximating communication in PNoCs. These results highlight the potential of using approximation strategies in PNoC architectures to reduce their energy footprint in emerging many-core platforms.

## ACKNOWLEDGEMENTS

This research was supported by the National Science Foundation (NSF) under grant number CCF-1813370.